\documentclass[12pt]{article}
\newtheorem{defn}{Definition}[section]

\newtheorem{thm}[defn]{Theorem}

\newtheorem{lemma}[defn]{Lemma}

\newtheorem{remark}[defn]{Remark}
\newtheorem{example}[defn]{Example}
\newtheorem{proc}[defn]{Procedure}
\newtheorem{cond}[defn]{Condition}

\newcommand{\brmk}{\begin{remark}\per\begin{em}}
\newcommand{\ermk}{\end{em}\end{remark}}
\newcommand{\bexa}{\begin{example}\per\begin{em}}
\newcommand{\eexa}{\end{em}\end{example}}
\newcommand{\bproc}{\begin{proc}\begin{em}}
\newcommand{\eproc}{\end{em}\end{proc}}

\newcommand{\beginsec}{\setcounter{equation}{0}}

\newcounter{bean}
\newcommand{\benuma}{\setlength{\labelwidth}{.25in}
\begin{list}
{(\alph{bean})}{\usecounter{bean}}}
\newcommand{\eenuma}{\end{list}}

\newcommand{\be}{\begin{equation}}
\newcommand{\ee}{\end{equation}}

\newcommand{\bea}{\begin{eqnarray}}
\newcommand{\eea}{\end{eqnarray}}

\newcommand{\beas}{\begin{eqnarray*}}
\newcommand{\eeas}{\end{eqnarray*}}

\newcommand{\goto}{\rightarrow}

\newcommand{\per}{\hspace{-.072in}{\bf .  }}

\newcommand{\lan}{\langle}
\newcommand{\ran}{\rangle}

\newcommand{\noi}{\noindent}

\newcommand{\skp}{\vspace{\baselineskip}}

\newcommand{\ink}{\rule{.5\baselineskip}{.55\baselineskip}}

\newcommand{\R}{I\!\!R}

\newcommand{\N}{{I\!\!N}}

\newcommand{\nin}{n \in \N}

\newcommand{\vphi}{\varphi}

\newcommand{\ital}{\textit}
\newcommand{\trm}{\textrm}

\newcommand{\lb}{\left\{}
\newcommand{\LL}{\ensuremath{\mathcal{L}}}

\newcommand{\lp}{\left(}

\newcommand{\rb}{\right\}}
\newcommand{\rp}{\right)}
\newcommand{\snsp}{\!}

\newcommand{\ve}{\ensuremath{\varepsilon}}

\newcommand{\X}{\ensuremath{\mathcal{X}}}

\newcommand{\Y}{\ensuremath{\mathcal{Y}}}

\newcommand{\wrq}{W_{n,r}}

\newcommand{\sqk}{S_{q,k}}
\newcommand{\drk}{D_{r,k}}
\newcommand{\idrk}{1_{\drk}(x)}
\newcommand{\ssum}{\sum_{k=1}^{\gamma_r}}
\newcommand{\Ltwo}{L^2(\Lambda,\theta)}

\renewcommand{\d}{\partial}

\begin{document}

\title{{\bf The Large Deviation Principle \\ for Coarse-Grained Processes}}
\author{Richard S. Ellis\thanks{This research was supported by a
grant from the Department of Energy (DE-FG02-99ER25376)
and by a grant from the National
Science Foundation (NSF-DMS-9700852).},
Kyle Haven\thanks{This research was supported by a
grant from the Department of Energy (DE-FG02-99ER25376).}, and
Bruce Turkington\thanks{This research was supported by a grant
from the Department of Energy (DE-FG02-99ER25376)
and by a grant from the National
Science Foundation (NSF-DMS-9971204).}\\
\small{rsellis@math.umass.edu, haven@math.umass.edu, turk@math.umass.edu}\\
Department of Mathematics and Statistics\\
University of Massachusetts\\
Amherst, MA  01003}

\maketitle

\thispagestyle{empty}

\begin{abstract}
The large deviation principle is proved for a class of 
$L^2$-valued processes that arise from the coarse-graining 
of a random field.  Coarse-grained processes of this kind 
form the basis of the analysis of local mean-field models 
in statistical mechanics by exploiting the long-range 
nature of the interaction function defining such models.  
In particular, the large deviation principle is used in a 
companion paper
\cite{EllHavTur2} 
to derive the variational principles that 
characterize equilibrium macrostates in statistical models 
of two-dimensional and quasi-geostrophic turbulence.  Such 
macrostates correspond to large-scale, long-lived flow 
structures, the description of which is the goal of the 
statistical equilibrium theory of turbulence.  The large 
deviation bounds for the coarse-grained process under 
consideration are shown to hold with respect to the strong 
$L^2$ topology, while the associated rate function is 
proved to have compact level sets with respect to the weak 
topology.  This compactness property is nevertheless 
sufficient to establish the existence of equilibrium 
macrostates for both the microcanonical and canonical 
ensembles.
\end{abstract}

\noi

\section{Introduction}
\label{intro}
\beginsec In many statistical mechanical models coarse-graining is a
fundamental construction that mediates between a microscopic scale on
which the model is defined and a macroscopic scale on which the model
is analyzed in a thermodynamic or continuum limit.  In this general
context, the coarse-graining is determined by an averaging procedure
on an intermediate scale.  Its usefulness relies on the property that
the functions defining the interactions of the model's variables can
be well approximated by corresponding functions of the averaged
variables.  Typically, this situation is met in models that have
long-range interactions and therefore have the character of local
mean-field theories.
For models of this kind, a complete and rigorous
analysis of the thermodynamic or continuum limit can be carried out
once the asymptotic behavior of an appropriate coarse-grained process
is characterized.  Such a characterization is provided by a large
deviation principle, which expresses in a sharp form the statistical
effect of the averaging procedure.

The well-known models of two-dimensional turbulence \cite{MilWeiCro,RobSom,Tur},
or more generally quasi-geostrophic turbulence \cite{DibMajTur,Hol},
are prime examples of this general class of local mean-field theories.
As we explain later, a coarse-grained process for these models is
defined by taking a certain local average of the underlying
microscopic vorticity field.  The dynamical invariants, which in these
models include the energy and circulation, are then expressed as
functions of this coarse-grained process via approximations that
become exact in the continuum limit.  With this representation in
hand, rigorous large deviation techniques allow one to deduce, in an
essentially intuitive way, the asymptotic behavior of the vorticity
field in the continuum limit.  In fluid dynamical terms, the large
deviation principle distinguishes certain mean flow structures, which
may take the form of jets or vortices, as the most probable
macrostates against a background of fluctuating, filamentary
microscopic vorticity.  In this way the large deviation analysis plays
a pivotal role in realizing the main goal of such models: to explain
the emergence and persistence of coherent structures within the
turbulent flow.

For the sake of definiteness, let us consider two-dimensional
turbulence in the unit torus $T^2 \doteq [0,1) \times [0,1)$.  The
underlying Hamiltonian system is governed by the Euler equations for
an inviscid, incompressible fluid with periodic boundary conditions on
the velocity and pressure fields.  This system is most conveniently
described as an evolution equation for the vorticity $\omega(x,t)
\doteq \d u_2 / \d x_1 - \d u_1 / \d x_2 $, which is the perpendicular
component of the curl of the velocity field $u=(u_1(x,t), u_2(x,t))$,
$\; x = (x_1,x_2) \in T^2 $.  With respect to this dynamics the vorticity
$\omega$ is advected, or rearranged, by the incompressible velocity
field $u$ that it induces instantaneously.  Generically, this
self-straining motion produces a fine-grained vorticity field $\omega$
that exhibits complex fluctuations on the small spatial scales.
Statistical equilibrium models are introduced to capture the essential
features of the flow without resolving the small-scale dynamics.

In order to construct a statistical equilibrium model, one discretizes
the dynamics and replaces the fine-grained vorticity field $\omega$ by
a microstate $\zeta$ defined on a suitable lattice $\LL_n$ in $T^2$.
Specifically, for each $\nin$ let $\LL_n$ be a uniform lattice of $a_n
\doteq 2^{2n}$ sites $s$ in $T^2$.  The intersite spacing in each
coordinate direction is $2^{-n}$.  Each such lattice of $a_n$ sites
induces a dyadic partition of $T^2$ into $a_n$ squares called
microcells, each having area $1/a_n$.  For each $s\in\LL_n$ we denote
by $M(s)$ the unique microcell containing the site $s$ in its lower
left corner.  The configuration spaces for the model are $\Y^{a_n}$,
where $\Y$ is a given closed subset of $\R$.  The elements of
$\Y^{a_n}$ are the microstates $\zeta = \{\zeta(s), s \in \LL_n\}$,
which we can identify with piecewise-constant vorticity fields $\zeta$
 relative to $\LL_n$; that is, $\zeta(x) = \zeta(s)$ for all $x
\in M(s)$, $s \in \LL_n$.

The probabilistic structure of the discretized microstate $\zeta$ is
chosen to be consistent with the postulated behavior of the
fine-grained vorticity field $\omega$. Specifically, it is determined
by a probability measure $P_n$ defined as follows.  Let $\rho$ be a
probability measure on $\R$ with support $\Y$.  If $\Y$ is unbounded,
assume that
\be
\label{eqn:finite}
\int_{\R} e^{\alpha y} \, \rho(dy) < \infty \ \mbox{ for all }
\alpha \in \R.
\ee
We then define
$P_n$ to be the product measure on $\Y^{a_n}$ with identical
one-dimensional marginals $\rho$.  With respect to $P_n$, the
collection $\{\zeta(s), s \in \LL_n\}$ is a finite family of
independent, identically distributed (i.i.d.) random variables having
common distribution $\rho$ and common range $\Y$.  Under a suitable
ergodic hypothesis, the given measure $\rho$ incorporates
the family of invariants associated with
incompressible rearrangement of vorticity.  Details are discussed in
\cite{BouEllTur2,EllHavTur2,Tur}.

The basic dynamical invariant of the Euler equations is the kinetic
energy, which is expressible as the following functional of the
vorticity:
\be \label{eqn:hamil}
H(\omega) \doteq \frac{1}{2} \int_{T^2 \times T^2} g(x-x') \,
\omega(x) \, \omega(x') \, dx dx'.
\ee
In this formula $g(x - x')$ is the generalized Green's function
for $-\bigtriangleup$ on $T^2$.
In the
lattice model on $\LL_n$, $H(\omega)$ is replaced by a lattice Hamiltonian
$H_n$.  This  is defined for $\zeta \in \Y^{a_n}$ by
\be
\label{eqn:mrham}
H_{n}(\zeta) \doteq \frac{1}{2 n^2} \sum_{s,s' \in \LL} g_n(s - s')
\zeta(s) \zeta(s') \, ,
\ee
where $g_n(s-s')$ is a certain lattice approximation to the
generalized Green's
function $g(x-x')$.  For instance,
$g_n(s-s')$ may be determined from a finite-difference discretization
or a spectral truncation.

The formalism of equilibrium statistical mechanics provides two joint
probability distributions for microstates $\zeta \in \Y^{a_n}$: the
microcanonical ensemble and the canonical ensemble.  In probabilistic
terms, the microcanonical ensemble expresses the conditioning of $P_n$
on the energy shell $\{\zeta \in \Y^{a_n} \! : \! H_n(\zeta) = E\}$,
where $E \in \R$ is a specified energy value.  In order to avoid
technical problems with the existence of regular conditional
probability distributions, we shall condition $P_n$ on the thickened
energy shell $\{H_n(\zeta) \in [E-\ve,E+\ve]\}$, where $\ve>0$.  Thus,
the microcanonical ensemble is the measure defined for Borel subsets
$B$ of $\Y^{a_n}$ by
\[
P_n^{E,\ve}\{B\}= P_n\{B \, | \, H_n \in [E-\ve,E+\ve]\} =
\frac{P_n\{B \cap \{H_n \in [E-\ve,E+\ve]\}\}}{P_n\{H_n \in [E-\ve,E+\ve]\}}.
\]
Correspondingly, the canonical ensemble is defined
for Borel subsets $B$ of $\Y^{a_n}$ by
\[
P_{n,\beta}\{B\}\doteq \frac{1}{Z(n,\beta)}
 \int_B \exp[-\beta H_n] \, dP_n.
\]
Here $\beta$ is a real number denoting the inverse temperature, and
$Z(n,\beta)$ is the partition function $\int_{\Y^{a_n}} \exp[-\beta
H_n] \, dP_n$, which normalizes the probability measure
$P_{n,\beta}$.

Models of this kind were originally proposed independently by Miller
et.\ al.\ \cite{Mil,MilWeiCro} and Robert et.\ al.\ \cite{Rob2,RobSom}
in the context of two-dimensional Euler flow.  Subsequently, they were
 extended to geophysical fluid dynamics, such as barotropic
quasi-geostrophic flow \cite{DibMajTur,EllHavTur2}. A model
of spin systems on a circle that exhibits an interesting phase
transition  has a similar, but simpler structure \cite{EisEll}.

In the setting of models of two-dimensional and quasi-geostrophic
turbulence, there are two basic goals of the equilibrium
statistical theory: first, to predict the formation of stable,
coherent flow structures from either the microcanonical ensemble or
the canonical ensemble; second, to deduce whether the two ensembles
yield equivalent results.  In order to achieve these two goals, which
depend on deriving properties of the two ensembles in the continuum limit,
 the crucial innovation is to introduce a
two-parameter stochastic process that is defined by a coarse-graining,
or local averaging, of the microscopic vorticity field over an
intermediate scale.  We now present this key construction.

Given $\nin$ and a positive integer $r < n$, we
consider a dyadic partition of the lattice $\LL_n$ into $\gamma_r
\doteq 2^{2r}$ blocks, each block containing $a_n/\gamma_r$ lattice
sites.  In correspondence with this partition we have a dyadic
partition $\{D_{r,k}, k =1,\ldots,\gamma_r\}$ of $T^2$ into
macrocells.  Each macrocell is the union of $a_n/\gamma_r$ microcells
$M(s)$.  This partition of $\LL_n$ into $\gamma_r$ blocks represents a
coarse-graining of the lattice $\LL_n$.  With respect to this partition,
we define the following coarse-grained process, obtained by
a local averaging over the sites of the macrocells $D_{r,k}$:
\be
\label{eqn:coarse-grain}
W_{n,r}(x) = W_{n,r}(\zeta,x) \doteq \sum_{k=1}^{\gamma_r} 1_{D_{r,k}}(x) \,
S_{n,r,k}(\zeta),
\ee
where
\be
\label{eqn:coarse-sum}
S_{n,r,k}(\zeta) \doteq \frac{1}{a_n/\gamma_r} \sum_{s \in D_{r,k}} \zeta(s).
\ee
The doubly indexed process $W_{n,r}$
takes values in the space $L^2(T^2)$.

The process $W_{n,r}$ has the following two properties, which will allow
us to evaluate its continuum limit with respect to either the microcanonical or canonical
ensemble.
\begin{enumerate}
\item In the double limit $n \goto \infty$, $r \goto \infty$,
with respect to the product measures $P_n$,
$W_{n,r}$ satisfies the large deviation principle (LDP) on $L^2(T^2)$
with scaling constants $a_n$ and an explicitly determined rate
function $I$.
\item In the double limit $n \goto \infty$, $r \goto \infty$, the
Hamiltonian $H_n(\zeta)$ is asymptotic to $H(W_{n,r}(\zeta))$
uniformly over microstates, where the functional
$H$ mapping $L^2(T^2)$ into $\R$
is defined in (\ref{eqn:hamil}); in symbols,
\begin{equation}  \label{eqn:hnapprox}
\lim_{r \goto \infty} \lim_{n \goto \infty} \sup_{\zeta \in \Y^{a_n}}
  |H_n(\zeta) - H(W_{n,r}(\zeta))| = 0.
\end{equation}
\end{enumerate}

The proof of the two-parameter LDP in item 1 is the main task of this
paper.  We will give a heuristic proof later in this section.  In
Section 2 we will formulate the LDP for a natural generalization of
$W_{n,r}$ and will prove this in Section 3.

The verification of (\ref{eqn:hnapprox}) in item 2 can be carried out
as in \cite[\S 4.2]{BouEllTur1}, where a similar approximation is
verified.  Essentially, this approximation depends on the fact that
the vortex interactions governed by $H_n$ are long-range, being
determined essentially by the Green's function $g(x-x')$.  For this
reason, $H_n$ is not sensitive to the small-scale structure of the
vorticity field and depends only on the local mean vorticity in the
continuum limit.  In other words, $H_n$ is well approximated by a
function of the coarse-grained process $W_{n,r}$.  This kind of
behavior is typical of local mean-field theories.  The turbulence
models under consideration here have the property that their local
mean-field approximations are asymptotically exact \cite{MilWeiCro}.

In the next part of this section, we motivate the two-parameter LDP in
item 1.  Later in Section 5 we indicate how items 1 and 2 together
allow one to evaluate the continuum limit of $W_{n,r}$ with
respect to the microcanonical
ensemble and the canonical ensemble.  These limits are expressed in
terms of variational formulas whose solutions correspond to coherent
structures for the two ensembles.  There we also discuss the question of
equivalence and nonequivalence of ensembles.

In order to motivate the LDP in item 1, we note that with respect to $P_n$
 the normalized sums $S_{n,r,k}$ are sample means of the
$a_n/\gamma_r$ i.i.d.\ random variables $\zeta(s), s \in D_{r,k}$.
Cram\'{e}r's Theorem therefore implies that, for each $k=1, \ldots \gamma_r$,
$\{S_{n,r,k}, n \in \N\}$ satisfies the
LDP with respect to $P_n$ with scaling constants $a_n/\gamma_r$ and with
the rate function
\[
i(z) \doteq \sup_{\alpha \in \R} \{\alpha z  - c(\alpha)\} \
\mbox{ for } z \in \R .
\]
This function is the
 lower semicontinuous, convex function conjugate to
the cumulant generating function
\[
c(\alpha) \doteq \log \int_{\R} e^{\alpha y}  \, \rho(dy),
\;\;\;\; \alpha \in \R \, .
\]
For $y_k \in \R$ we summarize the LDP for $S_{n,r,k}$
by the heuristic notation
\[
\lim_{n \goto \infty} \frac{1}{a_n/\gamma_r} \log P_n\{S_{n,r,k} \sim y_k\}
\approx - i(y_k).
\]

This basic LDP makes use of the fact that $c(\alpha)$ is finite for
all $\alpha \in \R$, which follows from the assumed property (\ref{eqn:finite})
of $\rho$.  The G\"{a}rtner-Ellis Theorem
allows one to extend Cram\'{e}r's Theorem to measures $\rho$ for which
$c(\alpha)$ is finite for $\alpha$ in a subset $A$ of $\R$ that
contains 0 in its interior and for which $\lim_{n \goto \infty}
|c\,'(\alpha_n)| = \infty$ whenever $\alpha_n$ is a sequence in
$\mbox{int}(A)$ converging to a boundary point of $\mbox{int}(A)$
\cite{DemZei}.  This extension is useful because it applies to measures $\rho$ having
exponential tails that arise in certain turbulence models
\cite{EllHavTur2}.

For each $\zeta$, $W_{n,r}(\zeta)$ is piecewise constant on the macrocells $D_{r,k}$.
To give a heuristic derivation of the LDP for
$W_{n,r}$, we approximate a
general $f \in L^2(T^2)$ by a piecewise constant function of the form
\[
\vphi(x) = \sum_{k=1}^{\gamma_r} \vphi_k \idrk.
\]
Then using Cram\'{e}r's Theorem for each $k=1, \ldots ,\gamma_r $
and the independence of $S_{n,r,1}, \ldots, S_{n,r,\gamma_r}$,
we have for all
sufficiently large $r$
\begin{eqnarray}
\label{eqn:bagel}
\lefteqn{
\lim_{n\rightarrow\infty}
          \frac{1}{a_n}\log P_n\!\lb W_{n,r}\sim f \rb}
\\ & & \approx  \lim_{n\rightarrow\infty}
          \frac{1}{a_n}\log P_n\!\lb W_{n,r}\sim \vphi \rb  \nonumber
\\  &  & = \frac{1}{\gamma_r}\lim_{n\rightarrow\infty}
        \frac{1}{a_n/\gamma_r}\log P_n\!\lb S_{n,r,1}\sim \vphi_1,
        \ldots S_{n,r,\gamma_r}\sim \vphi_{\gamma_r}\rb \nonumber \\
  & & =   \frac{1}{\gamma_r}\sum_{k=1}^{\gamma_r} \lim_{n\rightarrow\infty}
  \frac{1}{a_n/\gamma_r}\log P_n\!\lb S_{n,r,k} \sim \vphi_k\rb \nonumber
\\ & & \approx
- \frac{1}{\gamma_r}\sum_{k=1}^{\gamma_r} i(\vphi_k)
\; = \; - \int_{T^2} i(\vphi(x)) \, dx
  \; \approx \; - \int_{T^2} i(f(x)) \, dx. \nonumber
\end{eqnarray}
This calculation makes it reasonable to
expect that $W_{n,r}$ satisfies a two-parameter LDP
on $L^2(T^2)$
with the rate function
\begin{equation}  \label{eqn:rate_function}
I(f) \doteq \int_{T^2} i(f(x)) \, dx \, .
\end{equation}
Here and throughout the paper the term
``rate function on a space $\X$'' denotes a lower semicontinuous
function mapping $\X$ into $[0,\infty]$.  A rate function need not
have compact level sets.

In Section 2 we consider a natural generalization of the
doubly indexed process $W_{n,r}$ taking values in an $L^2$ space
and formulate an LDP in the strong topology on that space
[Thm.\ \ref{thm:ldp}].  The LDP is
proved in Section 3.  In
Section 4 we show that the rate function in this LDP has compact level sets
with respect to the weak topology on the $L^2$ space, though not in general
with respect to the strong topology.  The results in Sections 2, 3, and
4 are derived from basic principles using relatively straightforward
proofs.  These techniques are related to those introduced in
\cite{BouEllTur2}, which establishes an analogous two-parameter LDP
for a class of spatialized random measures.  Those random measures
also arise in the analysis of the continuum limit for turbulence
models \cite{BouEllTur1,MicRob}.  The results in the present paper, however,
are more elementary, both conceptually and technically.
At the end of Section 4 we comment on the paper \cite{BouEllTur2} in the light
of the present paper.  We also point out, in the next to last paragraph of
that section, an oversight in a proof in \cite{BouEllTur2}.

In Section 5 we summarize typical
applications of our main LDP stated in Theorem \ref{thm:ldp}.
In particular, we state the variational
principles for the microcanonical and canonical ensembles defined in this
introduction, and we demonstrate how to obtain these
principles from a large deviation analysis of the corresponding
ensembles.  The solutions of the variational principles
are called equilibrium macrostates.
In Section 5 we also point out how the existence of equilibrium
macrostates makes use of the weak compactness of the level sets of the
rate function.  With these results in hand, we then comment on the
equivalence and nonequivalence of ensembles for the microcanoncial and
canonical ensembles for which definitive results are given in
\cite{EllHavTur1}.  A complete discussion of the physical applications
of these results is contained in \cite{EllHavTur2}, where families
of stable, steady mean flows for a general class of geophysical fluid
dynamical models are characterized and computed.

\section{Statement of the LDP}
\label{sec:ldp}
\beginsec

In this section we formulate the LDP for a natural generalization of
the random functions $W_{n,r}$ defined
in (\ref{eqn:coarse-grain})--(\ref{eqn:coarse-sum}).
Let $(\Omega, {\cal F}, P)$ be a probability space and $d$ a positive
integer.  For each $r \in \N$, let $\{S_{n,r},n\in\N\}$ be a sequence
of random vectors mapping $\Omega$ into $\R^d$ and satisfying the LDP
as $n \goto \infty$ with positive scaling constants $c_{n,r}$ and convex
rate function $i$ independent of $r$.
In other words, for each $r$ $\, c_{n,r}
\goto \infty$ as $n \goto \infty$; for any
closed subset $F$ of $\R^d$
\[
\limsup_{n \goto \infty} \frac{1}{c_{n,r}} \log P\{S_{n,r} \in F\} \leq - i(F);
\]
and for any open subset $G$ of $\R^d$
\[
\liminf_{n \goto \infty} \frac{1}{c_{n,r}} \log P\{S_{n,r} \in G\} \geq - i(G) .
\]
Here $i(B)$ denotes the infimum of $i$ over the set $B$.  We assume
throughout that $i$ is lower semicontinuous and
convex on $\R^d$.  We do not assume
that $i$ has compact level sets, even though this
extra property is satisfied in many applications.  The setup in
Section 1 corresponds to choosing $S_{n,r}$ as in
(\ref{eqn:coarse-sum}) and $c_{n,r} = a_n/\gamma_r = 2^{2(n-r)}$
whenever $1 \leq r < n$; otherwise, $S_{n,r}$ and $c_{n,r}$ equal 0.

In order to give a general formulation of the LDP, we
consider a Polish space $\Lambda$ with metric $b$ (a complete separable
metric space) and let $\theta$ be a probability measure
on $\Lambda$.  $\Ltwo$ denotes the set of functions $f$ mapping $\Lambda$
into $\R^d$ and satisfying
\[
\|f\|^2_2 \doteq \int_{\Lambda} |f|^2 d\theta < \infty,
\]
where $|\cdot|$ denotes the Euclidean norm on $\R^d$.  Let $\gamma_r$
be a sequence of positive integers tending $\infty$.  For each
positive integer $n$ and $r$ we introduce $S_{n,r,1}, \ldots,
S_{n,r,\gamma_r}$, which are i.i.d.\ copies of $S_{n,r}$ mapping
$\Omega$ into $\R^d$.  For each $r \in \N$ we assume that $\Lambda$ is
partitioned into $\gamma_r$ subsets $D_{r,1},\ldots,D_{r,\gamma_r}$
and that these sets satisfy the following condition.

\begin{cond}\per  \label{part-cond}
For each $r\in\N$
 
\begin{enumerate}
  \item[{\em (i)}] $\theta\{D_{r,k}\}=1/\gamma_r$, $k = 1,\ldots,\gamma_r$,
  \item[{\em (ii)}] $\displaystyle\lim_{r \rightarrow\infty}
\max_{k\in\{1,\ldots,\gamma_r\}}
    \{\mbox{{\em diam}}(D_{r,k})\}=0$, where
$\mbox{{\em diam}}(D_{r,k}) \doteq \sup_{x,y \in D_{r,k}} b(x,y)$.
\end{enumerate}
 
\end{cond}
In Section 1 we worked with Lebesgue measure on
the unit torus $\Lambda=T^2$, which was partitioned
into $2^{2r}$ macrocells $\drk$.

The process whose asymptotics we wish to analyze is the doubly indexed
sequence of random functions defined for $\zeta \in \Omega$ and
$x \in \Lambda$ by
\begin{equation}  \label{eqn:wnr}
    W_{n,r}(x) = W_{n,r}(\zeta,x) \doteq
    \sum_{k=1}^{\gamma_r} 1_{D_{r,k}}(x) \, S_{n,r,k}(\zeta).
\end{equation}
Clearly, $W_{n,r}$
maps $\Omega$ into $\Ltwo$.  In Theorem \ref{thm:ldp} we formulate the
large deviation bounds for $W_{n,r}$ with respect to the strong
topology on that space.
The definition of $W_{n,r}$ in (\ref{eqn:wnr}) is more general than
in (\ref{eqn:coarse-grain}) because here we do not assume that
$S_{n,r,k}$ has the form (\ref{eqn:coarse-sum}).

The rate function that appears in the LDP for $W_{n,r}$ is defined next.
 
\begin{defn}\per \label{def:I}
Let $i$ denote the rate function in the LDP for $\{S_{n,r}, n \in \N\}$
on $\R^d$. Given $f \in \Ltwo$ define
\[  I(f) \doteq \int_{\Lambda}i \! \circ \! f \, d\theta.  \]
\end{defn}

Since $i$ is nonnegative and convex, it follows that $I$ is
well-defined, nonnegative and convex.  At the
end of Section 3 we prove that $I$
is lower semicontinuous with respect to the strong topology, which is
the topology generated by the open balls $B(f,\ve) \doteq \{g \in
\Ltwo : \|f-g\|_2 < \ve\}$ for $f \in \Ltwo$ and $\ve >0$. In general,
however, $I$ does not have compact level sets with respect to the
strong topology.  This is easily seen by returning to the setup of
Section 1, in which $S_{n,r,k}$ is a normalized sum of i.i.d.\ random
variables each distributed by $\rho$.  If we choose $\rho$ to be a
Gaussian measure on $\R$ having mean 0 and variance 1, then $i(z) =
\frac{1}{2}z^2$ and $I(f) = \frac{1}{2}\|f\|_2^2$.  In this case, level
sets of $I$ coincide with closed balls in $\Ltwo$ centered at the
origin, and these sets are not compact with respect to the strong
topology.  In Section 4 we prove, in a setting midway between those of
Sections 1 and 2,
that $I$ has compact level sets with respect to the weak
topology on $\Ltwo$ under the assumption that
$\rho$ decays at infinity at least as fast as a Gaussian.

We now state the two-parameter LDP for $W_{n,r}$, which
we prove in Section 3.

\begin{thm}\per  \label{thm:ldp}
We assume Condition {\em \ref{part-cond}}
and consider $\Ltwo$ with the strong topology.
Then the function $I$ given in Definition {\em \ref{def:I}}
maps $\Ltwo$ into $[0,\infty]$ and is lower semicontinuous, but in general
$I$ does not have compact level sets.  In addition, the
sequence $W_{n,r}$ satisfies the two-parameter
LDP on $\Ltwo$ with rate function $I$ in the
following sense.   For any strongly closed subset $F$
of $\Ltwo$
\[
  \limsup_{r \rightarrow\infty} \, \limsup_{n\rightarrow\infty}
    \frac{1}{\gamma_r c_{n,r}}\log P\snsp\lb W_{n,r}\in F\rb\leq
    -I(F),
\]
and for any strongly open subset $G$
of $\Ltwo$
\[
  \liminf_{r\rightarrow\infty} \, \liminf_{n\rightarrow\infty}
    \frac{1}{\gamma_r c_{n,r}}\log P\snsp\lb W_{n,r}\in G\rb\geq
    -I(G).
\]
\end{thm}

In our
companion paper \cite{EllHavTur2} we need the special case of Theorem \ref{thm:ldp}
discussed in Section 1, in which the coarse-grained process $W_{n,r}$ is defined by
(\ref{eqn:coarse-grain})--(\ref{eqn:coarse-sum}).


\section{Proof of Theorem \ref{thm:ldp}}
\label{proof}
\beginsec

In this section we prove the upper and lower large deviation bounds for
$W_{n,r}$ in separate, but elementary steps.
The following  lemma is used in both steps.

\begin{lemma} \per
\label{lem:sum-ldp}
For each $r$ the sequence $\{(S_{n,r,1},\ldots,S_{n,r,\gamma_r}), n \in \N\}$
satisfies the LDP on $(\R^d)^{\gamma_r}$ with scaling constants
$c_{n,r}$ and the rate function
\[
(\nu_1,\ldots,\nu_{\gamma_r}) \mapsto \sum_{k=1}^{\gamma_r} i(\nu_k).
\]
\end{lemma}

\noi {\bf Proof.}  The LDP is an immediate consequence of Lemmas
2.5--2.8 in \cite{LynSet}, since $S_{n,r,1}, \ldots, S_{n,r,\gamma_r}$
are i.i.d.\ copies of $S_{n,r}$ and each sequence $\{S_{n,r}, n \in
\N\}$ satisfies the LDP on $\R^d$ with scaling constants $c_{n,r}$ and
rate function $i$. \ink

\skp
\noi
{\it Proof of the Large Deviation Upper Bound}

Let $F$ be a strongly closed subset of $\Ltwo$.
For $r \in \N$ we define the closed set
\[
F_{r}  \doteq
\lb(\nu_1,\ldots,\nu_{\gamma_r}) \in (\R^d)^{\gamma_r}: \ssum \nu_k \idrk \in F \rb.
\]
We also define $L^2_r$ to be the set of $f \in \Ltwo$ of the form
$f(x) = \ssum \nu_k \idrk$ for some $\nu_1,\ldots, \nu_{\gamma_r}
\in \R^d$.
By Lemma \ref{lem:sum-ldp}, since $\theta\{\drk\}= 1/\gamma_r$,
\begin{eqnarray*}
\lefteqn{\limsup_{n\rightarrow\infty}\frac{1}{c_{n,r}}\log
  P\!\lb W_{n,r}\in F \rb}  \\
&&=\limsup_{n \goto \infty}  \frac{1}{c_{n,r}}\log P\!\lb(S_{n,r,1},\ldots,
S_{n,r,\gamma_r})\in F_{r}\rb \\
& & \leq  -\gamma_r \inf\lb\frac{1}{\gamma_r}\sum_{k=1}^{\gamma_r}i(\nu_{k}):
        (\nu_1,\ldots,\nu_{\gamma_r}) \in F_{r} \rb\\
  & & =  -\gamma_r \inf\lb\int_{\Lambda} i \! \circ \! f \, d\theta: f \in F \cap L^2_r \rb\\
  & & \leq  -\gamma_r \inf\lb\int_{\Lambda}i \! \circ \! f \, d\theta : f \in F \rb \\
  & & = -\gamma_r I(F).
\end{eqnarray*}
Dividing by $\gamma_r$ and sending $r \goto \infty$
gives the desired large deviation upper bound. \ink

\skp
\noi
{\it Proof of the Large Deviation Lower Bound}

In order to prove this bound, we need to approximate arbitrary
functions in $\Ltwo$ by functions that are piecewise-constant relative
to the partition $D_{r,k}, k=1, \ldots , \gamma_r$.  This is carried
out in the next lemma.

\begin{lemma} \per \label{lem:approx}
We assume Condition
{\em \ref{part-cond}}.  Let $f$ be any function in $\Ltwo$.
For $r \in \N$ define
\[
f^r(x) \doteq \sum_{k=1}^{\gamma_r} f_k^r \idrk, \ \mbox{ where } \
f_k^r \doteq  \gamma_r \! \int_{\drk} f \, d\theta.
\]
Then as $r \rightarrow \infty$, $\|f - f^r\|_2 \goto 0$.
\end{lemma}

\noi \textbf{Proof.}
For any given $\ve > 0$ there exists a bounded Lipschitz function
$\vphi$ mapping $\Lambda$ into $\R^d$ and satisfying $\|f - \vphi\|_2 < \ve$
\cite{Dud}.
Since the operator mapping $f \in \Ltwo
\mapsto f^r$ is an orthogonal projection,
\be
\label{eqn:frgr}
\|f^r - \vphi^r\|_2 = \|(f-\vphi)^r\|_2 \leq \|f-\vphi\|_2 < \ve.  \ee
Hence it suffices to estimate $\|\vphi - \vphi^r\|_2$, where $\vphi$ is a
Lipschitz function with constant $M < \infty$; that is, $|\vphi(x) - \vphi(y)|
\leq M b(y,x)$ for all $x, y \in \Lambda$.  Because the disjoint sets
$D_{r,k}$ have measure $\theta\{D_{r,k}\} = 1/\gamma_r$ [Cond.\
\ref{part-cond}(i)], a straightforward calculation gives
\begin{eqnarray*}
\|\vphi - \vphi^r\|_2^2 & = & \sum_{k=1}^{\gamma_r} \int_\Lambda 1_{D_{r,k}} |\vphi - \vphi^r_k|^2 \, d\theta \\
& = & \sum_{k=1}^{\gamma_r} \int_{D_{r,k}} \left|\gamma_r
\int_{D_{r,k}} [\vphi(y) - \vphi(x)] \, \theta(dx) \right|^2 \theta(dy) \\
& \leq & \sum_{k=1}^{\gamma_r} \int_{D_{r,k}} \left(\gamma_r
\int_{D_{r,k}} M \, b(y,x) \, \theta(dx) \right)^2 \theta(dy) \\
& \leq & \left(M \cdot \max_{k=1,\ldots,\gamma_r}
\{\mbox{diam}(D_{r,k})\}\right)^2.
\end{eqnarray*}
Sending $r \goto \infty$ and using Cond.\ \ref{part-cond}(ii), we
complete the proof. \ink

\skp
Given a strongly open subset $G$ of $\Ltwo$,
let $f$ be any function in $G$ and choose $\varepsilon >0$ so that
$B(f,\varepsilon) \subset G$.  Also choose $N \in \N$ such that for all
$r \geq N$ $B(f^r, \varepsilon/2) \subset B(f, \varepsilon)$, where $f^r$ is
the function defined in Lemma \ref{lem:approx}.  Such an $N$ exists
because of the $\Ltwo$-convergence of $f^r$ to $f$ proved in that lemma.
Define the open set
\[
G_{r,\varepsilon} \doteq \left\{ (\nu_1, \ldots , \nu_{\gamma_r}) \in
  (\R^d)^{\gamma_r}: \ssum \nu_k \idrk \in B(f^r,\ve/2) \right\}.
\]

Then for all $r \geq N$ Lemma \ref{lem:sum-ldp}  yields
\begin{eqnarray*}
 \lefteqn{\liminf_{n\rightarrow\infty}
          \frac{1}{\gamma_r c_{n,r}}\log P\!\lb W_{n,r}\in G\rb}  \\
&& \nonumber \geq \liminf_{n\rightarrow\infty}
          \frac{1}{\gamma_r c_{n,r}}
\log P\!\lb W_{n,r}\in B(f^r, \varepsilon/2)\rb  \\
&& = \liminf_{n\rightarrow\infty}
\frac{1}{\gamma_r c_{n,r}}\log P\!\lb (S_{n,r,1},\ldots,S_{n,r,\gamma_r})\in G_{r,\ve}\rb \nonumber \\
  & & \geq  -\frac{1}{\gamma_r}\inf\lb\sum_{k=1}^{\gamma_r}i(\nu_k):
                           (\nu_1,\ldots,\nu_{\gamma_r})\in G_{r,\ve}\rb  \nonumber \\
  & & \geq  -\frac{1}{\gamma_r}\sum_{k=1}^{\gamma_r} i(f^r_k)
\; = \;  -\frac{1}{\gamma_r}\sum_{k=1}^{\gamma_r}
                    i\!\lp \gamma_r \!\!\int_{D_{r,k}}
                               f \, d\theta \rp  \nonumber \\
  & & \geq  -\sum_{k=1}^{\gamma_r}\int_{D_{r,k}} i \! \circ \! f \, d\theta
  \; = \; -\int_{\Lambda} i \! \circ \! f \, d\theta
\; = \; -I(f). \nonumber
\end{eqnarray*}
The last inequality in this display follows from
Jensen's inequality.  The preceding display states that
\[
\liminf_{r \goto \infty} \, \liminf_{n\rightarrow\infty}
          \frac{1}{\gamma_r c_{n,r}}\log P\!\lb W_{n,r}\in G\rb \geq  - I(f)
\]
for arbitrary $f \in G$.  Taking the supremum of
$-I(f)$ over $f \in G$ yields the desired large deviation lower
bound.  \ink
 
\skp
\noi
{\it Proof that $I$ Is Strongly Lower Semicontinuous}

We end this section by proving that $I$ is lower semicontinuous with
respect to the strong topology on $\Ltwo$.  Namely, we show that
$\liminf_{n \rightarrow \infty} I(f_n) \geq I(f)$ for any strongly
convergent sequence $f_n \goto f$ in $\Ltwo$.

There exists a subsequence $\{f_{n_k}\}$ such
that $\lim_{n_k \goto \infty} I(f_{n_k}) = \liminf_{n \goto \infty}
I(f_n)$ and $f_{n_k} \goto f$ $\theta$-a.s.  Fatou's lemma and the
lower semicontinuity of $i$ on $\R^d$ then yield
\begin{eqnarray*}
\liminf_{n \rightarrow \infty} I(f_n) &=& \liminf_{n_k \goto \infty}
I(f_{n_k}) \: = \: \liminf_{n_k \rightarrow \infty} \int_{\Lambda} i
\! \circ \! f_{n_k} \, d\theta \\ & \geq & \int_{\Lambda} \liminf_{n_k
\rightarrow \infty} i \! \circ \! f_{n_k} \, d\theta \: \geq \:
\int_{\Lambda} i \! \circ \! f \, d\theta \: = \: I(f).
\end{eqnarray*}
This completes the proof of strong lower semicontinuity.    \ink


\section{Weak versus Strong Topology}
\label{sec:compact}
\beginsec

In the first half of this section we prove, in the general setting of
Section 2, that the function $I$ in Definition \ref{def:I} is lower
semicontinuous with respect to the weak topology on $\Ltwo$ [Thm.\
\ref{thm:weaklsc}].  We also show, in a setting midway between
those of Sections 1 and 2,
that $I$ has compact level sets with
respect to the weak topology [Thm.\ \ref{thm:weakcompact}].

The fact that $I$ is weakly lower semicontinuous
and has weakly compact level sets is used to establish the existence of
equilibrium macrostates for both the microcanonical and canonical ensembles
introduced in Section 1.  We recall that in Section 1 $\Lambda$
equals $T^2$ and $\theta$ equals Lebesgue measure.
In the microcanonical case the equilibrium macrostates are
characterized as solutions of the following constrained
minimization problem: minimize $I(f)$ over $f \in L^2(T^2)$
subject to $H(f) \in [ E - \ve,E + \ve ]$. Direct methods in the
calculus of variations then assure that a minimizer exists since
the functional $H$ is weakly continuous on $L^2(T^2)$.
Similarly, in the canonical case the equilibrium macrostates are
characterized as solutions of the following minimization problem:
 minimize $I(f) + \beta H(f)$ over $f \in L^2(T^2)$. Again, a
minimizer exists by virtue of the properties of $I$ and $H$ with
respect to the weak topology on $L^2(T^2)$.  These applications
are further discussed in Section 5 and in \cite{EllHavTur2}.

A related issue arises in the Miller-Robert theory of
coherent structures in two-dimensional turbulence \cite{MilWeiCro,Rob2}.
In that theory the generalized enstrophy invariants
\[
A(\omega) \doteq \int_{\Lambda} a(\omega) \, d\theta
\]
are included together with the energy invariant $H$.  A family of
moment functions $a$ parameterize these extra constraints; these
functions may be chosen arbitrarily provided certain regularity and growth
conditions are satisfied \cite{BouEllTur2,Tur}.   Unlike the Hamiltonian $H$, the
functionals $A$ are generally not continuous with respect to the
weak topology on $\Ltwo$.  The classical, quadratic enstrophy
$A(\omega)=
\int_{\Lambda} \frac{1}{2}\omega^2 \, d\theta$ provides a
counterexample. Moreover, as the same example shows, the crucial
approximation property (\ref{eqn:hnapprox}) of $H$ is not shared
by the functionals $A$. For this reason, in the case of the
Miller-Robert theory one must rely on a coarse-graining process
at the level of empirical measures rather than at the level of
sample means.  Details are given in \cite{BouEllTur1,BouEllTur2}.
After the proof of Theorem \ref{thm:weakcompact}, we comment on
the connection between the main results in the present paper and
those in \cite{BouEllTur2} and then point out an oversight in a
proof in \cite{BouEllTur2}.

\skp
\noi
{\it $I$ Is  Weakly Lower Semicontinuous}

We work in the general setting of Section 2.
Let $\lan \cdot,\cdot \ran$ denote the inner product on $\Ltwo$.
The weak topology on $\Ltwo$ is generated by the neighborhoods
\[
B(f; f_1,\ldots,f_p, \ve) \doteq \{g \in \Ltwo:
|\lan f,f_i\ran_2 - \lan g,f_i\ran_2| < \ve, i=1,\ldots,p\}
\]
for $f \in \Ltwo, p \in \N,
f_1,\ldots,f_p \in \Ltwo$, and $\ve > 0$.

\begin{thm}\per
\label{thm:weaklsc}
$I$ is lower semicontinuous with respect to the weak topology
on $\Ltwo$.
\end{thm}

\noi {\bf Proof.} Our strategy is to approximate $I$ by a sequence of
functionals for which the lower semicontinuity is almost immediate.
For $r \in \N$ and $f \in \Ltwo$ define $I^r(f) \doteq I(f^r)$, where
$f^r$ is the approximating function given in Lemma \ref{lem:approx}.
We claim that $I(f) = \sup_{r \in \N} I^r(f)$.  On the one hand,
Jensen's inequality ensures that
\[
I^r(f) = \int_\Lambda i \! \circ \! f^r \, d\theta =
\frac{1}{\gamma_r} \sum_{k=1}^{\gamma_r} i\!\left(\gamma_r
\int_{D_{r,k}} f \, d\theta\right) \; \leq \, I(f).
\]
On the other hand, since $f^r \goto f$ strongly in $\Ltwo$ [Lem.\
\ref{lem:approx}], the lower limit
\[
\label{eqn:b}
\liminf_{r \goto \infty} I^r(f) \geq I(f)
\]
is valid by the strong lower semicontinuity of $I$ proved at the end of Section
3.  Hence if we prove that each $I^r$ is weakly lower
semicontinuous, the weak lower semicontinuity of $I$ follows.

Since the sum of weakly lower semicontinuous functions is weakly lower
semicontinuous, it suffices to prove that for each $k$ the function
mapping $f \mapsto i(\gamma_r \int_{D_{r,k}} f \, d\theta)$ is weakly
lower semicontinuous.  This is an immediate consequence of the facts
that the linear mapping $f \mapsto \gamma_r \int_{D_{r,k}} f \,
d\theta$ is weakly continuous and that the extended real-valued function $i$ is
lower semicontinuous.   \ink

\skp
\noi {\it $I$ Has Weakly Compact Level Sets}

We carry out this analysis in a setting generalizes that of Section 1,
but is more special than that of Section 2.  Let $\Lambda$ be a Polish
space, $\theta$ a probability measure on $\Lambda$, and $a_n$ and
$\gamma_r$ two sequences of positive integers tending to $\infty$.
Consider a subset $\LL_n$ of $\Lambda$ consisting of $a_n$ points.
For each $r$, $\Lambda$ is assumed to be partitioned into $\gamma_r$
sets $D_{r,1},\ldots,D_{r,\gamma_r}$ satisfying Condition
\ref{part-cond}.  We then define $W_{n,r}$ as in
(\ref{eqn:coarse-grain})--(\ref{eqn:coarse-sum}).  We also
impose the additional condition that the measure $\rho$ that defines
the common distribution of $\zeta(s), s \in \LL_n$, decays at infinity
at least as fast as a Gaussian.

\begin{thm}\per
\label{thm:weakcompact}
We assume that there exists $\delta > 0$ such that
\be
\label{eqn:decay}
\int_{\R^d} \exp\!\left(\frac{\delta}{2} |y|^2\right) \rho(dy) < \infty.
\ee
Then the level sets of $I$ are compact with respect to the weak topology
on $\Ltwo$.
\end{thm}

\noi
{\bf Proof.}
First, we claim that
there exists $D \in (0,\infty)$ such that for all $z \in \R^d$
\[
i(z) \geq \frac{\delta}{2}  |z|^2 - D.
\]
The inequality $\lan \alpha,y \ran \leq \frac{1}{2\delta}|\alpha|^2 +
\frac{\delta}{2} |y|^2$ for
$\alpha$ and $y$ in $\R^d$ implies that
\[
c(\alpha) \leq \frac{|\alpha|^2}{2\delta} +
\log \int_{\R^d} \exp\!\left(\frac{\delta}{2}
|y|^2\right) \rho(dy) =  \frac{|\alpha|^2}{2\delta} + D.
\]
Hence
\[
i(z) \doteq \sup_{\alpha \in \R^d} \{\lan \alpha,z \ran - c(\alpha)\}
\geq \sup_{\alpha \in \R^d} \left\{\lan \alpha,z \ran -
\frac{|\alpha|^2}{2\delta}\right\} - D
= \frac{\delta}{2} |z|^2 - D.
\]
This establishes the claim.

Using this estimate, we see that for any $f \in \Ltwo$
\begin{equation} \label{eqn:Igrows}
I(f) \doteq \int_{\Lambda} i \! \circ \! f \, d\theta \geq
\frac{\delta}{2} \|f\|_2^2 - D,
\end{equation}
from which it follows that for $M < \infty$
\[
\{f \in \Ltwo : I(f) \leq M\} \subset \left\{f \in \Ltwo : \|f\|_2^2 \leq
\frac{2}{\delta} (M + D)\right\}.
\]
Since the closed balls
$\{f \in \Ltwo : \|f\|_2^2 \leq \frac{2}{\delta} (M + D)\}$ are weakly compact,
the weak compactness of the level sets of $I$ follows from the fact
that the level sets are weakly closed.  This is a consequence of the weak
lower semicontinuity of $I$ proved in Theorem \ref{thm:weaklsc} \ink

\skp
\noi {\it Comments on Coarse-Grained Empirical Measures}

Here we comment briefly on the paper \cite{BouEllTur2} in the
light of the present work.  In that paper we proved the LDP for a
general class of doubly indexed
random measures.  For the purpose of comparison with the present
paper, we consider only a special case of those random measures
having a form analogous to
(\ref{eqn:coarse-grain})--(\ref{eqn:coarse-sum}).
In a notation analogous to that in the present paper, define
the coarse-grained empirical measures
\begin{equation} \label{cgem}
\tilde{W}_{n,r}(dx \times dy ) =
\tilde{W}_{n,r}(\zeta, dx \times dy ) \doteq \theta(dx) \otimes
   \sum_{k=1}^{\gamma_r} 1_{D_{r,k}}(x) \, L_{n,r,k}(\zeta, dy),
\ee
where
\be
\label{eqn:em}
L_{n,r,k}(\zeta,dy) \doteq \frac{1}{a_n/\gamma_r} \sum_{s \in D_{r,k}}
\delta_{\zeta(s)}(dy).
\ee
The LDP for  $\tilde{W}_{n,r}$ follows
from Sanov's Theorem \cite{DemZei}, just as the LDP for  $W_{n,r}$
in the present paper
follows from Cramer's Theorem.  The rate function  associated with
$\tilde{W}_{n,r}$ is the relative entropy $R( \cdot | \theta \times \rho)$.
As is well-known, the relative entropy is weakly lower semicontinuous
and has compact level sets with respect to the weak topology on
the space of probability measures
on $\Lambda \times \Y$.

The doubly indexed process $W_{n,r}(\zeta,x)$ considered in the present paper
equals the density, with respect to $\theta(dx)$, of the mean with
respect to $dy$ of $\tilde{W}_{n,r}(\zeta,dx \times dy)$.  However, since
the mapping taking a measure in two variables to the density, with
respect to $\theta(dx)$, of the mean with respect to $dy$ is not
continuous, it is not efficient to try to prove the LDP for $W_{n,r}$
from that for $\tilde{W}_{n,r}$.  Instead, it is much better to construct
the self-contained proof of the LDP for $W_{n,r}$ given in Section 3
of the present paper, using the ideas introduced in \cite{BouEllTur2}
for the analysis of $\tilde{W}_{n,r}$.

An abstract setting analogous to the setup in Section 2 of this paper
is introduced in Section 2 of \cite{BouEllTur2}.  In that generality,
the purported proof in \cite{BouEllTur2} that the rate function $J$
given in Definition 2.3 has compact level sets involves the following
circular reasoning (see pages 318--319).  While $\mu$ in the last
display on page 318 depends on $r$, the $r$ appearing in the first
display on page 319 depends on $N$, which in turn depends on $\mu$.
The conclusion is that the proof is invalid.

Although we cannot conclude in general that the quantity $J$ in
\cite{BouEllTur2} has compact level sets, this need not be a serious
hindrance.  Indeed, in many cases one can prove directly from the form
of $J$ that it has compact level sets; a number of examples are given
in Example 2.7 in \cite{BouEllTur2}.  The most basic of these examples
is given in part (a), where the rate function equals the relative
entropy; in this case the compactness of the level sets is automatic.
It is this particular example that is used in the applications paper
\cite{BouEllTur1}.

\section{Discussion of Applications}
\label{sec:appl}
\beginsec

In this final section we return to the setting of Section 1 and
indicate how Theorem \ref{thm:ldp} is applied to characterize
equilibrium macrostates for the turbulence models.  As in Section 1, we
consider both the microcanoncial ensemble and the canonical ensemble, taking into
account the energy constraint.  After summarizing
 how Theorems \ref{thm:weaklsc} and \ref{thm:weakcompact} are used to
prove the existence of equilibrium macrostates,
we point out another interesting property of these macrostates;
namely, the stability of the steady mean flows that the macrostates determine.

The theoretical tools needed to carry out this analysis
 are fully developed
in \cite{EllHavTur1} for a general class of statistical equilibrium
models of local mean-field type.  In that paper we derive, from underlying
LDP's, variational
principles for the equilibrium macrostates in both the microcanonical and
canonical ensembles.  Moreover, we give complete
and definitive results concerning the equivalance of these ensembles
at the level of their equilibrium macrostates, emphasizing the possibility
of nonequivalence.

In an applied companion paper \cite{EllHavTur2} these general results
are applied to a widely used geophysical model; namely, barotropic,
quasi-geostrophic turbulence in a zonal channel.  The results in
\cite{EllHavTur2} rely on the LDP stated in Theorem \ref{thm:ldp}
in the present paper for the coarse-grained process
$W_{n,r}$.  For the geophysical model we find that
nonequivalence of ensembles occurs over a wide range of the physical
parameters.  This surprising
result is then shown to be related to stability
conditions on the steady mean flows determined by the equilibrium
macrostates.  The main conclusions of the paper are that when nonequivalence
prevails, equilibrium macrostates corresponding to the microcanonical
ensemble are richer than those corresponding to the canonical ensemble
and furthermore that the microcanonical equilibrium macrostates
express the essential features of the
coherent structures that form in geophysical fluid flows.

We now summarize some of these results, stressing their connections to
the probabilistic questions addressed in the preceding sections of
this paper.

\skp
\noi
{\it Variational Principles for Equilibrium Macrostates}

We first consider the microcanonical ensemble $P_n^{E,\ve}$,
where $E$ is an admissible energy value; i.e., one for which
the constraint set $\{f \in L^2(T^2): H(f) = E\}$ is nonempty.
For admissible $E$, we prove in Theorem 3.2 of \cite{EllHavTur1}
that the coarse-grained vorticity
field $W_{n,r}$ satisfies the LDP on $L^2(T^2)$
with scaling constants $a_n \doteq 2^{2n}$ in the
continuum limit $n \goto \infty$, $r \goto \infty$, $\ve \goto 0$.
The rate function $I^E$ for this LDP is given explicitly in terms of
the rate function $I(f)$ defined in (\ref{eqn:rate_function}); namely,
\be
\label{eqn:microrate}
I^E(f) \doteq \left\{ \begin{array}{ll} I(f) + S(E) & \trm{if }\;
H(f) = E, \\ \infty & \trm{otherwise,}
\end{array} \right.
\ee
where
\be
\label{eqn:entropy}
S(E) \doteq - \inf_{g \in L^2(T^2)} \{I(g) : H(g) = E\}.
\ee
The quantity $S(E)$ is called the microcanonical entropy.
The LDP with respect to the microcanonical ensemble
follows readily from the LDP, given in Theorem \ref{thm:ldp},
for $W_{n,r}$ with respect to the product measures $P_n$.

For $f \in L^2(T^2)$ satisfying $H(f) = E$, we summarize the LDP
with respect to $P_n^{E,\ve}$ by the formal asymptotic statement
\be
\label{eqn:microheur}
P_n^{E,\ve}\{W_{n,r} \sim f\} \approx \exp[-a_n I^E(f)] \ \mbox{ as }
n \goto \infty, r \goto \infty, \ve \goto 0.
\ee
In this setting, $L^2(T^2)$ is the state space of the coarse-grained
process, and its elements $f$ are the macrostates or coarse-grained
vorticity fields.

The set of microcanonical equilibrium macrostates is defined to be
\[
{\cal E}^E \doteq \{f \in L^2(T^2) : I^E(f) = 0\} .
\]
This set plays a central role in the theory.
For any $f \in L^2(T^2) \setminus {\cal E}^E$
we have $I^E(f) > 0$. The formal statement (\ref{eqn:microheur}) suggests
that the macrostate $f$ has
an exponentially small probability of being observed as a
coarse-grained vorticity field in the continuum limit of the microcanonical
ensemble.  As a consequence, (\ref{eqn:microheur}) suggests, and the
 LDP for $W_{n,r}$ with respect to the microcanonical ensemble
 allows one to prove, that the macrostates $f \in {\cal E}^E$ are the
overwhelmingly most probable coarse-grained vorticity fields compatible with the
microcanonical constraint $H(f) = E$.  For this reason, the
macrostates in ${\cal E}^E$ determine long-lived, large-scale coherent
structures in the turbulent vorticity field, the prediction of which
is the goal of the statistical equilibrium theory.

Analogous results apply to the canonical ensemble $P_{n,\beta}$ for
any $\beta \in \R$.  In order to obtain the LDP, the inverse temperature $\beta$
must be scaled with $a_n$; the physical reason for this is given in
\cite{BouEllTur1,Tur}.
With respect to $P_{n,a_n \beta}$, the same coarse-grained process
$W_{n,r}$ satisfies the LDP with scaling constants $a_n$ in the
continuum limit $n \goto \infty$, $r \goto \infty$ \cite[Thm.\ 2.4]{EllHavTur1}.
The rate function
$I_\beta$ is given by
\be
\label{eqn:canonrate}
I_\beta(f) \doteq I(f) + \beta H(f) - \varphi(\beta),
\ee
where
\[
\varphi(\beta) \doteq \inf_{g \in L^2(T^2)} \{I(g) + \beta H(g)\}.
\]
Again, this LDP for $W_{n,r}$ follows readily from the LDP given in Theorem
\ref{thm:ldp}.

The set of canonical equilibrium macrostates is defined by
\[
{\cal E}_\beta \doteq \{f \in L^2(T^2) : I_\beta(f) = 0\}.
\]
 The relationship between the set ${\cal E}^E$ and
the microcanonical measures $P_n^{E,\ve}$ is mirrored by the relationship
between ${\cal E}_\beta$ and the scaled canonical measures $P_{n,a_n \beta}$.
Namely, with respect to the latter measures, ${\cal E}_\beta$
consists of the overwhelmingly most probable coarse-grained states.
They correspond to
long-lived, large-scale coherent structures within a turbulent
vorticity field at a given $\beta$.  It is difficult, however, to
justify on physical grounds prescribing a ``turbulent temperature''
$1/\beta$, especially when $\beta <0$.  This negative temperature
regime is nevertheless the one of most physical interest in real
applications.

\skp
\noi
{\it Existence of Equilibrium Macrostates}

Throughout this discussion of the existence of equilibrium
macrostates, we assume that $\rho$ satisfies the decay condition
given in (\ref{eqn:decay}).
We first ask whether there exist microcanonical equilibrium macrostates
for each admissible energy value $E$.  In other words, for each admissible
$E$ is the set ${\cal E}^E$ nonempty?
Because of (\ref{eqn:microrate}), determining the equilibrium macrostates in ${\cal E}^E$ is
equivalent to solving the constrained minimization problem
\[
\mbox{ minimize } I(f) \mbox{ over }
          \{f \in L^2(T^2) : H(f) = E \} .
\]
Analogously, we ask whether there exist canonical equilibrium
macrostates for each given inverse temperature $\beta$.  In other
words, is the set ${\cal E}_{\beta}$ nonempty?
Because of (\ref{eqn:canonrate}), determining the
equilibrium macrostates in ${\cal E}_{\beta}$ is
equivalent to solving the unconstrained minimization problem
\[
\mbox{ minimize } I(f) + \beta H(f) \mbox{ over }
         f \in L^2(T^2) .
\]
These variational problems are dual in the sense that the
Lagrange multiplier for the constraint $H(f)=E$ in the microcanonical
problem is the prescribed parameter $\beta$ in the canonical
problem.

In both variational problems, the existence of solutions is assured by
Theorems \ref{thm:weaklsc} and \ref{thm:weakcompact}.  With respect to
the weak topology on $L^2(T^2)$,
$I$ is lower semicontinuous, its level sets are compact,
and $H$ is continuous.    Consequently, the direct methods
of the calculus of variations apply to the microcanonical and canonical
problems, yielding the existence of minimizers in both cases.

The following mean-field equation is satisfied by a solution $f$ of
either the microcanonical or canonical variational principle:
\begin{equation} \label{eqn:mfe}
i'(f) \,=\, -\beta \int_{T^2} g(x-x') f(x') \, dx'  \, ,
\end{equation}
where, as in (\ref{eqn:hamil}), $ g(x-x')$ is the generalized Green's function.
This equation shows that the most probable, coarse-grained vorticity
fields $f$ in both ensembles correspond to steady,
deterministic flows \cite{BouEllTur1,Tur}.
These first-order conditions are identical for the two ensembles, except
that $\beta$ is specified in the canonical ensemble while $\beta$ is
determined along with the solution $f$ in the microcanonical ensemble.

\skp
\noi
{\it Stability of Equilibrium Macrostates}

Even though the mean-field equations satisfied by equilibria in ${\cal
E}^E$ and in ${\cal E}_{\beta}$ are identical, the correspondence
between these sets of minimizers is subtle.
This issue is commonly
called the equivalence of ensembles; it investigates the relationships
between the set of solutions of the constrained minimization problem
that characterizes microcanonical equilibrium macrostates and the set
of solutions of the unconstrained minimization problem that
characterizes canonical equilibrium macrostates.  This topic is
discussed in great detail in \cite{EllHavTur1} for a wide range of
models that includes as a special case the model of barotropic
quasi-geostrophic turbulence studied in \cite{EllHavTur2}.

As we show in \cite{EllHavTur1},
the equivalence or nonequivalence of ensembles depends entirely on concavity
properties of the microcanonical entropy $S$, which is defined in
(\ref{eqn:entropy}).
In general, the microcanonical equilibrium
macrostates are richer than the canonical equilibrium macrostates.
This assertion is a consequence of the following results.
Every $f \in {\cal E}_\beta$ lies in ${\cal E}^E$ for some $E$.
If $S$ is strictly concave and smooth, then the usual
thermodynamic relation $\beta = \partial S/\partial E$ defines a
one-to-one correspondence between the two families of equilibrium
sets.  If, on the other hand, $S$ is not concave at a value
$E = E^*$, then the
ensembles are nonequivalent in the sense that ${\cal E}^{E^*}$ is disjoint
from the sets ${\cal E}_\beta$ for all values of $\beta$. A precise
and general formulation of this striking behavior is given in Theorem
4.4 in \cite{EllHavTur1}.  Moreover, in Section 2 of \cite{EllHavTur1}
and in Section 6 of \cite{EllHavTur2}, a number of examples of models
of turbulence are given in which the microcanonical entropy is not
concave over a substantial subset of its domain, and so the ensembles
are nonequivalent.

The question of equivalence between the microcanonical and canonical
ensembles is intimately related to stability conditions for the
equilibrium macrostates $f$.  Such stability criteria are derived from the
second-order conditions satisfied by a minimizer $f$, and these
conditions are different for the constrained and unconstrained
variational problems.
With respect to the canonical ensemble, the functional $I +
\beta H$ itself provides a Lyapunov functional at $f$ whenever
$f$ is a nondegenerate minimizer.  This construction, which is known
as Arnold nonlinear stability analysis in the deterministic context,
proves that a perturbation of a canonical equilibrium macrostate $f$ that
is small in the $L^2$ norm remains close to $f$ in the $L^2$ norm for all
time.  Interestingly, the strong $L^2$ topology for this stability
theorem coincides with the topology for which the LDP holds in the
statistical equilibrium theory.  In this setting, the LDP can be viewed as a
weak form of a stability statement for microstates; under an ergodic evolution
of the microstates, the coarse-grained process remains close to the
equilibrium macrostate in the $L^2$ norm.

The familiar Arnold construction, however, is not adequate to prove
the stability of microcanonical equilibrium macrostates when the ensembles
are not equivalent.  Nevertheless, a refined argument based on
penalizing the functional $I + \beta H$ furnishes the needed Lyapunov
functional for the microcanonical ensemble.  Then an $L^2$ stability
result analogous to the one mentioned for the canonical ensemble is valid.
This refined stability analysis fills an important gap in the known
stability criteria for two-dimensional flows and their geophysical
counterparts.  The reader is referred to \cite{EllHavTur2} for a
complete discussion.

\skp
\noi
{\it Summary}

Section 5 of this paper shows the importance, in applications, of
Theorem \ref{thm:ldp}, which proves the
LDP for the doubly indexed process $W_{n,r}$ with respect to the product
measures $P_n$.  The special
case of this process given in
(\ref{eqn:coarse-grain})--(\ref{eqn:coarse-sum}) is needed
in our companion paper \cite{EllHavTur2}, which
gives a rather complete analysis of the equilibrium macrostates for a basic
geophysical model.  For this model Theorem \ref{thm:ldp} allows one to prove LDP's
for the coarse-grained process $W_{n,r}$ with respect to both the microcanonical
ensemble and the canonical ensemble.  In turn, these LDP's allow one
to characterize equilibrium macrostates with respect to both ensembles, the
equivalence and nonequivalence of which are determined by concavity properties
of the microcanonical entropy.  In addition, in \cite{EllHavTur2} stability properties
of the equilibrium macrostates are derived, using both the familiar Arnold
construction and an extension of this construction based on penalizing the
Lyapunov functional used in the Arnold construction.
The fundamental innovation in this work is coarse-graining, which, via
 the doubly indexed process $W_{n,r}$, allows one to mediate between a
microscopic scale on which the model is defined and a macroscopic scale on which
the equilibrium macrostates are defined and analyzed.  The LDP for $W_{n,r}$
given in Theorem \ref{thm:ldp}
is the basic mathematical result that makes all the other analysis possible.


\end{document}